\documentclass[runningheads]{llncs}
\usepackage{amssymb}
\usepackage{graphicx}
\usepackage{llncsdoc}
\usepackage{makeidx, epsfig,amsmath}  
\usepackage{url}
\usepackage{makecell}

\usepackage[colorinlistoftodos]{todonotes}
\usepackage[colorlinks=true, allcolors=blue]{hyperref}
\usepackage{caption}
\usepackage[title]{appendix}
\usepackage{amsfonts}
\usepackage{multirow}
\usepackage{booktabs}
\usepackage[ruled,vlined,algo2e]{algorithm2e}

\usepackage{adjustbox}
\usepackage[colorinlistoftodos]{todonotes}
\usepackage[colorlinks=true, allcolors=blue]{hyperref}
\usepackage{epsfig}
\usepackage{graphicx}
\usepackage{amsmath}
\usepackage{amssymb}
\usepackage{algorithm}
\usepackage{algorithmic}
\usepackage{subfigure}
\usepackage{caption}
\usepackage{epsfig}
\usepackage{epstopdf}
\usepackage{wrapfig}
\usepackage{multirow}
\usepackage{array}
\usepackage{tabulary}
\usepackage{graphicx}
\usepackage{color}
\usepackage{marvosym}

\makeatletter\def\Hy@Warning#1{}\makeatother

\newcolumntype{P}[1]{>{\centering\arraybackslash}p{#1}}

\newcolumntype{K}[1]{>{\centering\arraybackslash}p{#1}}
\newcolumntype{L}[1]{>{\raggedright\arraybackslash}p{#1}}
\newcolumntype{C}[1]{>{\centering\arraybackslash}p{#1}}
\newcolumntype{R}[1]{>{\raggedleft\arraybackslash}p{#1}}

\usepackage[font=small,labelfont=bf]{caption}

\newcommand{\Section}[1]%

\begin{document}
%
%

\title{Improved Prognostic Prediction of Pancreatic Cancer Using Multi-Phase CT by Integrating Neural Distance and Texture-Aware Transformer}
\author{Hexin Dong\inst{1,2,3},\thanks{Work was done during an internship at Alibaba DAMO Academy. Corresponding authors$^{(\textrm{\Letter})}$: \href{mailto: yaojiawen.yjw@alibaba-inc.com}{yaojiawen.yjw@alibaba-inc.com}, \href{mailto:zhangli_pku@pku.edu.cn}{zhangli\_pku@pku.edu.cn}, \href{mailto:18940259980@163.com}{18940259980@163.com}.}  Jiawen Yao\inst{1,3}$^{(\textrm{\Letter})}$, Yuxing Tang\inst{1}, Mingze Yuan\inst{1,2,3}, Yingda Xia\inst{1}, Jian Zhou\inst{4,5}, Hong Lu\inst{6}, Jingren Zhou\inst{1,3}, Bin Dong\inst{2,7}, Le Lu\inst{1}, Zaiyi Liu\inst{8}, Li Zhang\inst{2}$^{(\textrm{\Letter})}$, Yu Shi\inst{9}$^{(\textrm{\Letter})}$ and  Ling Zhang\inst{1} }
\authorrunning{H. Dong et al.}
\titlerunning{Improved Prognostic Prediction of Pancreatic Cancer}

\institute{$^1$DAMO Academy, Alibaba Group
$^2$Peking University \\
$^3$ Hupan Lab, 310023, Hangzhou, China \\
$^4$Sun Yat-sen University Cancer Center 
$^5$South China Hospital, Shenzhen University
$^6$Tianjin Medical University Cancer Institute and Hospital\\
$^7$Peking University Changsha Institute for Computing and Digital Economy
$^8$Guangdong Provincial People’s Hospital
$^9$Shengjing Hospital
}


\maketitle              

\begin{abstract}
Pancreatic ductal adenocarcinoma (PDAC) is a highly lethal cancer in which the tumor-vascular involvement greatly affects the resectability and, thus, overall survival of patients. However, current prognostic prediction methods fail to explicitly and accurately investigate relationships between the tumor and nearby important vessels. 
This paper proposes a novel learnable neural distance that describes the precise relationship between the tumor and vessels in CT images of different patients, adopting it as a major feature for prognosis prediction. Besides, 
different from existing models that used CNNs or LSTMs to exploit tumor enhancement patterns on dynamic contrast-enhanced CT imaging, we improved the extraction of dynamic tumor-related texture features in multi-phase contrast-enhanced CT by fusing local and global features using CNN and transformer modules, further enhancing the features extracted across multi-phase CT images. 
We extensively evaluated and compared the proposed method with existing methods in the multi-center (n=4) dataset with 1,070 patients with PDAC, and statistical analysis confirmed its clinical effectiveness in the external test set consisting of three centers. The developed risk marker was the strongest predictor of overall survival among preoperative factors and it has the potential to be combined with established clinical factors to select patients at higher risk who might benefit from neoadjuvant therapy. 

\end{abstract}


\section{Introduction}
\noindent 
Pancreatic ductal adenocarcinoma (PDAC) is one of the deadliest forms of human cancer, with a 5-year survival rate of only 9\%~\cite{siegel2019cancer}. Neoadjuvant chemotherapy can increase the likelihood of achieving a margin-negative resection and avoid unnecessary surgery in patients with aggressive tumor types\cite{Yuan_2023_CVPR}. Providing accurate and objective preoperative biomarkers is crucial for triaging patients who are most likely to benefit from neoadjuvant chemotherapy. However, current clinical markers such as larger tumor size and high carbohydrate antigen (CA) 19-9 level may not be sufficient to accurately tailor neoadjuvant treatment for patients~\cite{tsai2020importance}. Therefore, multi-phase contrast-enhanced CT has a great potential to enable personalized prognostic prediction for PDAC, leveraging its ability to provide a wealth of texture information that can aid in the development of accurate and effective prognostic models~\cite{koay2019computed,BianRadiology2023}.

Previous studies have utilized image texture analysis with hand-crafted features to predict the survival of patients with PDACs~\cite{attiyeh2018survival}, but the representational power of these features may be limited. In recent years, deep learning-based methods have shown promising results in prognosis models~\cite{lou2019image,ChengYao2021CCR,feng2022end}. However, PDACs differ significantly from the tumors in these studies. A clinical investigation based on contrast-enhanced CT has revealed a dynamic correlation between the internal stromal fractions of PDACs and their surrounding vasculature~\cite{prokesch2002isoattenuating}. Therefore, focusing solely on the texture information of the tumor itself may not be effective for the prognostic prediction of PDAC. It is necessary to incorporate tumor-vascular involvement into the feature extraction process of the prognostic model. Although some studies have investigated tumor-vascular relationships~\cite{YAO2021102150,yao2022deep}, these methods may not be sufficiently capable of capturing the complex dynamics between the tumor and its environment.

\begin{figure*}[!htb]
	\centering
	\includegraphics[width=0.8\linewidth]{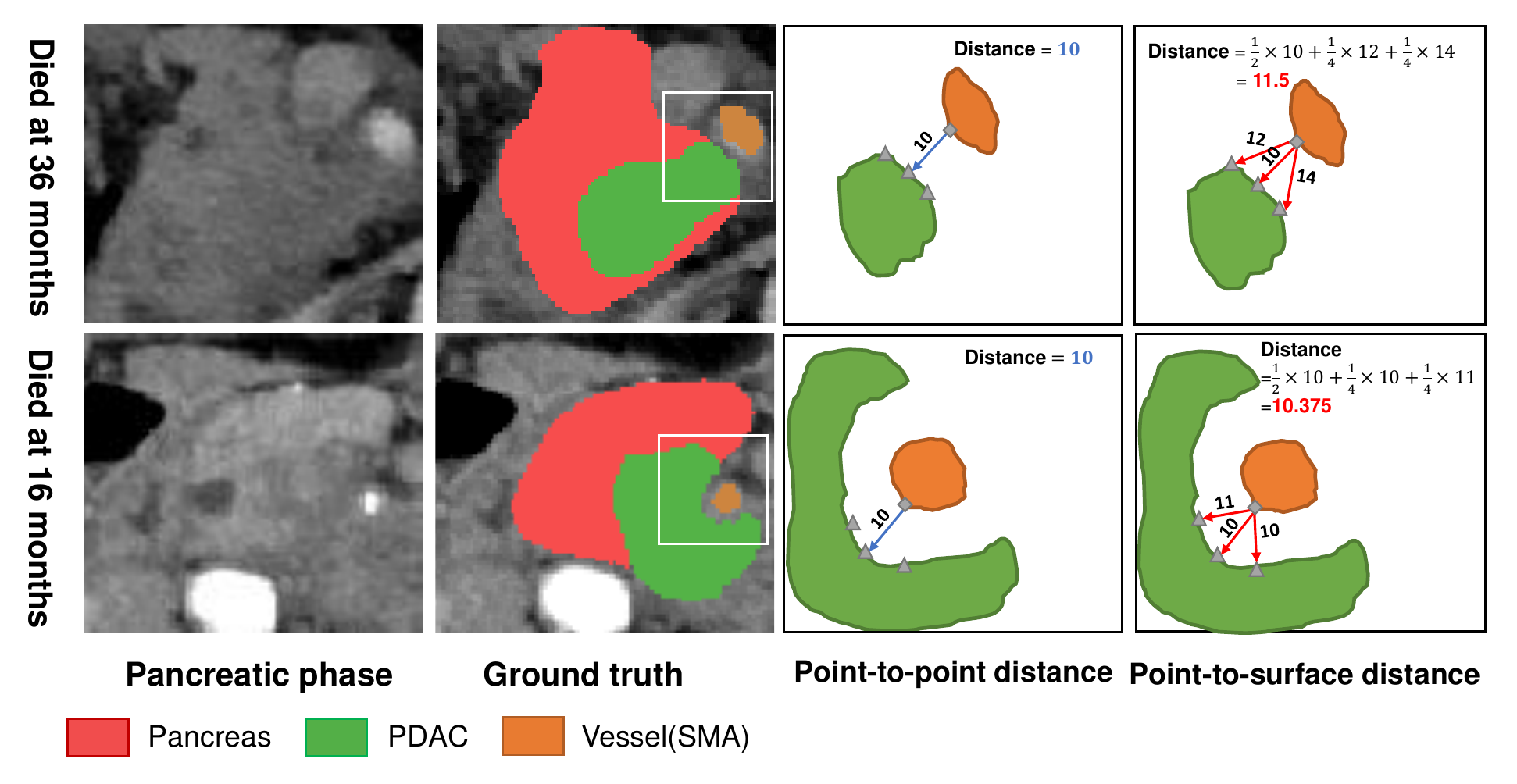}
	\caption{Two examples of spatial information between vessel (orange region) and tumor (green region). The minimum distance, which refers to the closest distance between the Superior Mesenteric Artery (SMA) and the PDAC tumor region, is almost identical in these two cases. We define the surface-to-surface distance based on point-to-surface distance (weighted-average of red lines from $\diamondsuit$ to $\triangle$) instead of point-to-point distance (blue lines) to better capture the relationship between the tumor and the perivascular tissue. Here $\diamondsuit$ and $\triangle$ are points sampled from subset $\hat{\mathcal{V}_c}$ and $\hat{\mathcal{P}_c}$ defined in \autoref{equ:close}. The distances and weights shown in the figure is for illustration purposes only.}
\label{fig:dis} 
\end{figure*}

We propose a novel approach for measuring the relative position relationship between the tumor and the vessel by explicitly using the distance between them. Typically, Chamfer distance~\cite{Chamfer}, Hausdorff distance~\cite{2002Comparing}, or other surface-awareness metrics are used. However, as shown in \autoref{fig:dis}, these point-to-point distances cannot differentiate the degree of tumor-vascular invasion~\cite{Pancreatic}. To address this limitation, we propose a learnable neural distance that considers all relevant points on different surfaces and uses an attention mechanism to compute a combined distance that is more suitable for determining the degree of invasion. Furthermore, to capture the tumor enhancement patterns across multi-phase CT images, 
we are the first to combine convolutional neural networks (CNN) and transformer~\cite{dosovitskiy2020vit} modules for extracting the dynamic texture patterns of PDAC and its surroundings. This approach takes advantage of the visual transformer's adeptness in capturing long-distance information compared to the CNN-only-based framework in the original approach. By incorporating texture information between PDAC, pancreas, and peripancreatic vessels, as well as the local tumor information captured by CNN, we aim to improve the accuracy of our prognostic prediction model.

In this study, we make the following contributions:
(1) We propose a novel approach for aiding survival prediction in PDAC by introducing a learnable neural distance that explicitly evaluates the degree of vascular invasion between the tumor and its surrounding vessels.
(2) We introduce a texture-aware transformer block to enhance the feature extraction approach, combining local and global information for comprehensive texture information. We validate that the cross-attention is utilized to capture cross-modality information and integrate it with in-modality information, resulting in a more accurate and robust prognostic prediction model for PDAC.
(3) Through extensive evaluation and statistical analysis, we demonstrate the effectiveness of our proposed method. The signature built from our model remains statistically significant in multivariable analysis after adjusting for established clinical predictors. Our proposed model has the potential to be used in combination with clinical factors for risk stratification and treatment decisions for patients with PDAC.

\section{Methods}
\begin{figure*}[t]
	\centering
	\includegraphics[width=0.95\linewidth]{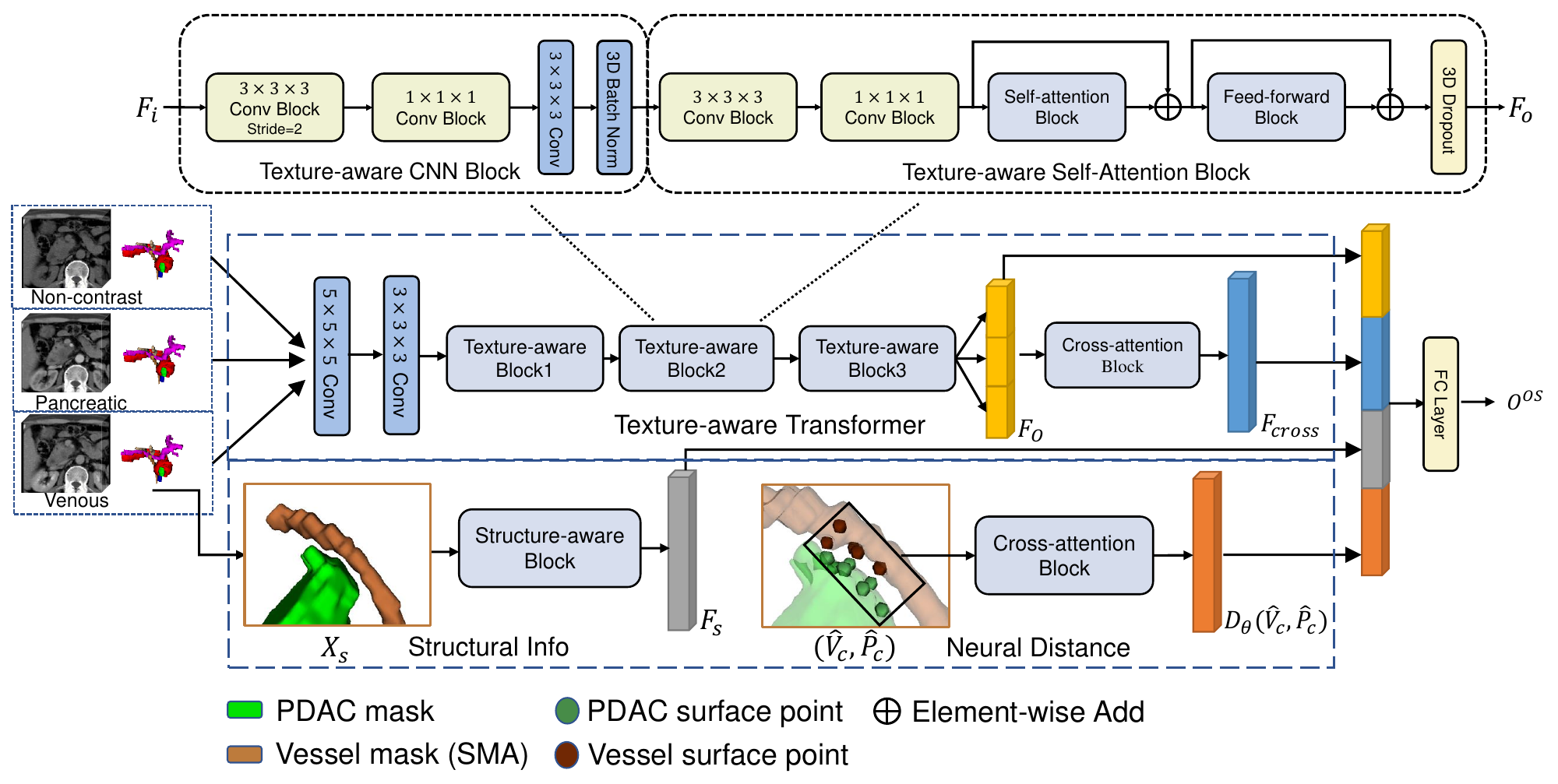}
	\caption{An overview of the proposed method. The texture-aware transformer captures texture information among PDAC, Pancreas and vessels around Pancreas with our proposed texture-aware transformer block and a cross-attention block to fusion cross-modality features. The structure-aware block extracts the structure relationship between PDAC and four related vessels. The neural distance calculates the distances between the PDAC surface and the vessel surface with our proposed neural distance. We first select related points set from the closest sub-surface on PDAC and vessels respectively. Then we use a cross-attention block to obtain the neural distance. Finally, we concatenate features from three branches to obtain the survival outcome $O_{OS}$. }
	\label{fig: Framework} 
\end{figure*}

As shown in \autoref{fig: Framework}, the proposed method consists of two main components. The first component combines the CNN and transformer to enhance the extraction of tumor dynamic texture features. The second component proposes a neural distance metric between PDAC and important vessels to assess their involvements.

\subsection{Texture-Aware Vision Transformer: Combination of CNN and Transformer}
\label{sec:texture}
Recently, self-attention models, specifically vision transformers (ViTs~\cite{dosovitskiy2020vit}), have emerged as an alternative to CNNs in survival prediction~\cite{zheng2022multi,saeed2022tmss}. Our proposed texture-aware transformer, inspired by MobileViT~\cite{mehta2021mobilevit}, aims to combine both local information (such as PDAC texture) and global information (such as the relationship between PDAC and the pancreas). This approach is different from previous methods that rely solely on either CNN-based or transformer-based backbones, focusing only on local or global information, respectively.

The texture-aware transformer (\autoref{fig: Framework}) comprises three blocks, each consisting of a texture-aware CNN block and a texture-aware self-attention block. These blocks encode the input feature of an image $\mathbf{F}_i \in \mathbb{R}^{H\times W \times D \times C}$ to the hidden feature $\mathbf{F}_c \in \mathbb{R}^{H\times W \times D \times C_{l}}$ using a $3\times3\times3$ convolutional layer, followed by a $1\times1\times1$ convolutional layer. The $3\times3\times3$ convolution captures local spatial information, while the $1\times1\times1$ convolution maps the input tensor to a higher-dimensional space (\textit{i.e.}, $C_l>C$). The texture-aware CNN block downsamples the input, and the texture-aware self-attention block captures long-range non-local dependencies through a patch-wise self-attention mechanism.

In the texture-aware self-attention block, the input feature $\mathbf{F}_c$ is divided into N non-overlapping 3D patches $\mathbf{F}_u \in \mathbb{R}^{V\times N \times C_u}$, where $V = hwd$ and $N = HWD/V$ is the number of patches, and $h,w,d$ are the height, width, and depth of a patch, respectively. For each voxel position within a patch, we apply a multi-head self-attention block and a feed-forward block following~\cite{NIPS2017_3f5ee243} to obtain the output feature $\mathbf{F}_o$.
In this study, preoperative multi-phase CE-CT pancreatic imaging includes the non-contrast phase, the  pancreatic phase and venous phase. Therefore, we obtain three outputs from the transformer block with the input of these phases, denoted as $\mathbf{F}_o^1, \mathbf{F}_o^2, \mathbf{F}_o^3 \in \mathbb{R}^{D\times C}$, resulting in the concatenated output $\mathbf{F}_o \in \mathbb{R}^{D \times 3C}$.

Instead of directly fusing the outputs as in previous work, we employ a 3-way cross-attention block to extract cross-modality information from these phases. The cross-attention is performed on the concatenated self-attention matrix with an extra mask $\mathbf{M} \in \{0,-\infty\}^{3C \times 3C}$, defined as:
\begin{equation}
    \begin{aligned}
     \mathbf{F}_{cross} &= 
     \operatorname{Softmax}(\mathbf{Q}\mathbf{K}^{\mathrm{T}}+ \mathbf{M})\mathbf{V},\\
     \mathbf{M}(i,j) &= 
        \begin{cases}
            -\infty & kC < i,j \leq (k+1)C,\quad k =0,1,2,\\
            0 & \text{otherwise,}
    
        \end{cases}        
    \end{aligned}    
    \label{eq1}
\end{equation}
Here, $\mathbf{Q},\mathbf{K},\mathbf{V}$ are the query, key, and value matrices, respectively, obtained by linearly projecting the input $\mathbf{F}_{o}^{\mathrm{T}} \in \mathbb{R}^{3C \times D}$. The cross-modality output $ \mathbf{F}_{cross}$ and in-modality output $\mathbf{F}_o^{\mathrm{T}}$ are then concatenated and passed through an average pooling layer to obtain the final output feature of the texture branch, denoted as $\mathbf{F}_t \in \mathbb{R}^{C_{t}}$.

\subsection{Neural Distance: Positional and Structural Information between PDAC and Vessels}

The vascular involvement in patients with PDAC affects the resectability and treatment planning~\cite{ducreux2015cancer}. In this study, we investigate four important vessels: portal vein and splenic vein (PVSV), superior mesenteric artery (SMA), superior mesenteric vein (SMV), and truncus coeliacus (TC). We used a semi-supervised nnUnet model to segment PDAC and the surrounding vessels, following recent work~\cite{yao2022deep,koehler2023noisy}. We define a general distance between the surface boundaries of PDAC ($\mathcal{P}$) and the aforementioned four types of vessels ($\mathcal{V}$) as $D(\mathcal{V},\mathcal{P})$, which can be derived as follows:
\begin{equation}
  D(\mathcal{V},\mathcal{P})=d_{ss}(\mathcal{V},\mathcal{P})+d_{ss}(\mathcal{P},\mathcal{V})=\frac{1}{\Vert \mathcal{V} \Vert}\int_{\mathcal{V}}d_{ps}(v,\mathcal{P})\mathrm{d}v+\frac{1}{\Vert \mathcal{P} \Vert}\int_{\mathcal{P}}d_{ps}(p,\mathcal{V})\mathrm{d}p,
\end{equation}
where $v\in \mathcal{V}$ and $p\in \mathcal{P}$ are points on the surfaces of blood vessels and PDAC, respectively. The point-to-surface distance $d_{ps}(v,\mathcal{P})$ is the distance from a point $v$ on $\mathcal{V}$ to $\mathcal{P}$, defined as $d_{ps}(v,\mathcal{P})= \min_{p\in \mathcal{P}}\;\Vert v-p \Vert_2^2$, and vice versa.

To numerically calculate the integrals in the previous equation, we uniformly sample from the surfaces $\mathcal{V}$ and $\mathcal{P}$ to obtain the sets $\hat{\mathcal{V}}$ and $\hat{\mathcal{P}}$ consisting of $N_v$ points and $N_p$ points, respectively. The distance is then calculated between the two sets using the following equation:
\begin{equation}
\label{equ:set_dis}
  D(\hat{\mathcal{V}},\hat{\mathcal{P}})=\frac{1}{N_v}\sum_{v \in \hat{\mathcal{V}}} d_{ps}(v,\hat{\mathcal{P}})+\frac{1}{N_p}\sum_{p \in \hat{\mathcal{P}}} d_{ps}(p,\hat{\mathcal{V}}).  
\end{equation}

However, the above distance treats all points equally and may not be flexible enough to adapt to individualized prognostic predictions. Therefore, we improve the above equation in two ways. Firstly, we focus on the sub-sets $\hat{\mathcal{V}}_c$ and $\hat{\mathcal{P}}_c$ of $\hat{\mathcal{V}}$ and $\hat{\mathcal{P}}$, respectively, which only contain the K closest points to the opposite surfaces $\hat{\mathcal{P}}$ and $\hat{\mathcal{V}}$, respectively. The sub-sets are defined as:
\begin{align}
\begin{aligned}
\label{equ:close}
    \hat{\mathcal{V}}_c &= \operatorname{argmin}_{\{v_1, v_2, \cdots, v_K\} \subset \hat{\mathcal{V}} } \sum_{i=1}^{K} d_{ps}(v_i, \hat{\mathcal{P}}),\\
    \hat{\mathcal{P}}_c &= \operatorname{argmin}_{\{p_1, p_2, \cdots, p_K\} \subset \hat{\mathcal{P}} } \sum_{i=1}^{K} d_{ps}(p_i, \hat{\mathcal{V}}).
\end{aligned}
\end{align}

\noindent Secondly, we regard the entire sets $\hat{\mathcal{V}}_c$ and $\hat{\mathcal{P}}_c$ as sequences and calculate the distance using a 2-way cross-attention block (similar to \autoref{eq1}) to build a neural distance based on the 3D spatial coordinates of each point:
\begin{equation}\label{equ:4}
    D_{\mathbf{\theta}}(\hat{\mathcal{V}},\hat{\mathcal{P}})=\operatorname{CrossAttention}(\hat{\mathcal{V}}_c,\hat{\mathcal{P}}_c), \quad \hat{\mathcal{V}}_c,\hat{\mathcal{P}}_c\in \mathbb{R}^{K\times 3}.
\end{equation}
Neural distance allows for the flexible assignment of weights to different points and is able to find positional information that is more suitable for PDAC prognosis prediction.
In addition to neural distance, we use the 3D-CNN model introduced in \cite{YAO2021102150} to extract the structural relationship between PDAC and the vessels. Specifically, we concatenate each PDAC-vessel pair $\mathbf{X}^{v}_{s}\in \mathbb{R}^{2\times H \times W \times D}$, where $v\in$\{PVSV, SMV, SMA, TC\} and obtain the structure feature $\mathbf{F}_{s} \in \mathbb{R}^{C_{s}}$.

Finally, we concatenate the features extracted from the two components and apply a fully-connected layer to predict the survival outcome, denoted as $O^{OS}$, which is a value between 0 and 1. To optimize the proposed model, we use the negative log partial likelihood as the survival loss~\cite{katzman2018deepsurv}.

\section{Experiments}
\textbf{Dataset.} In this study, we used data from Shengjing Hospital to train our method with 892 patients, and data from three other centers, including Guangdong General Hospital, Tianjin Medical University and Sun Yat-sen University Cancer Center for independent testing with 178 patients. The contrast-enhanced CT protocol included non-contrast, pancreatic, and portal venous phases. PDAC masks for 340 patients were manually labeled by a radiologist from Shengjing Hospital with 18 years of experience in pancreatic cancer, while the rest were predicted using self-learning models~\cite{koehler2023noisy,zhang2020robust} and checked by the same annotator. Other vessel masks were generated using the same semi-supervised segmentation models.
\textbf{C-index} was used as our primary evaluation metric for survival prediction. We also reported the survival \textbf{AUC}, which estimates the cumulative area under the ROC curve for the first 36 months.

\noindent\textbf{Implementation details:} We used nested 5-fold cross-validation and augmented the training data by rotating volumetric tumors in the axial direction and randomly selecting cropped regions with random shifts. We also set the output feature dimensions to $C_t=64$ for the texture-aware transformer, $C_{s}=64$ for the structure extraction and $K=32$ for the neural distance. The batch size was 16 and the maximum iteration was set to 1000 epochs, and we selected the model with the best performance on the validation set during training for testing. We implemented our experiments using PyTorch 1.11 and trained the models on a single NVIDIA 32G-V100 GPU.

\noindent\textbf{Ablation Study.} We first evaluated the performance of our proposed texture-aware transformer (TAT) by comparing it with the ResNet18 CNN backbone and ViT transformer backbone, as shown in \autoref{tab:ablation}. Our model leverages the strengths of both local and global information in the pancreas and achieved the best result. Next, we compared different methods for multi-phase stages, including LSTM, early fusion (Fusion), and cross-attention (Cross) in our method. Cross-attention is more effective and lightweight than LSTM. Moreover, we separated texture features into in-phase features and cross-phase features, which is more reasonable than early fusion.

 Secondly, we evaluated each component in our proposed method, as shown in \autoref{fig: Framework}, and presented the results in \autoref{tab:ablation}. Combining the texture-aware transformer and regular structure information improved the results from 0.630 to 0.648, as tumor invasion strongly affects the survival of PDAC patients. We also employed a simple 4-variable regression model that used only the Chamfer distance of the tumor and the four vessels for prognostic prediction. The resulting C-index of 0.611 confirmed the correlation of the distance with the survival, which is consistent with clinical findings~\cite{Pancreatic}. Explicitly adding the distance measure further improved the results. Our proposed neural distance metric outperformed traditional surface distance metrics like Chamfer distance, indicating its suitability for distinguishing the severity of PDAC.

\begin{table}[t]\caption{Ablation tests with different network backbones including ResNet18(Res), ViT and texture-aware transformer(TAT) and methods for multi-phases including LSTM, early fusion(Fusion) and cross-attention(Cross).}
\centering
\begin{tabular}{|c|c|c|c|c|} \hline

 Network Backbone  & Structural Info & Distance & Model Size(M) & C-index \\ \hline
Res-LSTM & - & - &  65.06 & $0.618 \pm 0.017$ \\ \hline
Res-Cross  & - & - & 43.54 & $0.625\pm 0.016$ \\ \hline
ViT-Cross  & - & - & 23.18 & $0.628\pm 0.018$\\ \hline
TAT-Fusion & - & - & \textbf{3.64} & $0.626\pm0.022$\\ \hline
TAT-Cross  & - & - & 15.13& $0.630\pm0.019$\\ \hline
TAT-Cross  & \checkmark & - & 15.90 &  $0.648\pm 0.021$ \\ \hline
-  & - & Chamfer distance & - & $0.611 \pm 0.029$\\ \hline
TAT-Cross  & \checkmark & Chamfer distance & 15.93 & $0.652 \pm 0.019$\\ \hline
TAT-Cross  & \checkmark & Nerual distance & 16.28 & $\textbf{0.656} \pm 0.017$ \\ \hline
\end{tabular}
\label{tab:ablation}
\end{table}

\begin{table}[t]\caption{Results of different methods on nested 5-fold CV and independent set.}
\centering
\begin{adjustbox}{width=1\textwidth}
\begin{tabular}{l|c|c|P{2cm}|c}
\hline
  & \multicolumn{2}{c|}{Nested 5-fold CV(n=892)} & \multicolumn{2}{c}{Independent test(n=178)}     \\ \hline
         & C-index   & AUC   & C-index  & AUC\\ \hline
{3DCNN-P~\cite{lou2019image}} & {0.630 $\pm$ 0.009}   & 0.668 $\pm$ 0.019 & {0.674}    & 0.740 \\ \hline
{Early Fusion~\cite{tang2020deep}} &{0.635 $\pm$ 0.011}   & 0.670 $\pm$ 0.024 & {0.696}    & 0.779 \\ \hline
DeepCT-PDAC~\cite{yao2022deep} & 0.640 $\pm$ 0.018  & 0.680 $\pm$ 0.036 & {0.697}    & 0.773 \\ \hline
{Ours}        & {$\textbf{0.656} \pm 0.017$}   & \textbf{0.695} $\pm$ 0.023 & {\textbf{0.710}}    & \textbf{0.792} \\
 \hline
 \end{tabular}
 \end{adjustbox}
 \label{tab:baseline}
\end{table}

\noindent\textbf{Comparisons.} To further evaluate the performance of our proposed model, we compared it with recent deep prediction methods~\cite{tang2020deep,yao2022deep} and report the results in \autoref{tab:baseline}. We modified baseline deep learning models~\cite{lou2019image,tang2020deep} and used their network architectures to take a single pancreatic phase or all three phases as inputs. DeepCT-PDAC~\cite{yao2022deep} is the most recent method that considers both tumor-related and tumor-vascular relationships using 3D CNNs. Our proposed method, which uses the transformer and structure-aware blocks to capture tumor enhancement patterns and tumor-vascular involvement, demonstrated its effectiveness with better performance in both nested 5-fold cross-validation and the multi-center independent test set.

In \autoref{tab:stats}, we used univariate and multivariate Cox proportional-hazards models to evaluate our signature and other clinicopathologic factors in the independent test set. The proposed risk stratification was a significant prognostic factor, along with other factors like pathological TNM stages. After selecting significant variables ($p<0.05$) in univariate analysis, our proposed staging remained strong in multivariable analysis after adjusting for important prognostic markers like pT and resection margins. Notably, our proposed marker remained the strongest among all pre-operative markers, such as tumor size and CA 19-9.

\begin{table}[t]\caption{Univariate and Multivariate Cox regression analysis. HR: hazard ratio.}
\centering
\begin{tabular}{l|c|c|c|c}
\hline
  Independent test set (n=178)  & \multicolumn{2}{l|}{Univariate Analysis} & \multicolumn{2}{l}{Multivariate Analysis}     \\ \hline
         & HR (95\% CI)   & p-value   & HR (95\% CI)               & p-value \\ \hline
Proposed (High vs low risk)    &  2.42(1.64-3.58)  &   $<$0.0001 &    1.85(1.08-3.17)              & 0.027       \\ \hline
Age ($> 60$ vs $\leq 60$)     &  1.49(1.01-2.20)  &   0.043 &    1.01(0.65-1.58)              & 0.888       \\ \hline
Sex (Male vs Female)     &  1.28(0.86-1.90)    & 0.221 &  -  &-      \\ \hline
pT (pT3-pT4 vs pT1-pT2)         &   3.17(2.10-4.77)  & $<$0.0001  &   2.44(1.54-3.86)     &  0.00015      \\ \hline
 pN (Positive ve Negative) &  1.47(0.98-2.20)      &    0.008  & 1.34(0.85-2.12)   &0.210  \\ \hline
Resection margin (R1 vs R0) &  2.84(1.64-4.93)   &    $<$0.0001 &1.68(0.92-3.07) &0.091     \\ \hline
 CA19-9 ($> 210 $ vs $\leq210$ U/mL)     &  0.94(0.64-1.39)      &    0.759  &   -  &-    \\ \hline
Tumor Size ($> 25 $ vs $\leq25$ mm)  & 2.36(1.59-3.52)  &   $<$0.0001  & 0.99(0.52-1.85)&0.963\\ \hline
Tumor Location (Head vs Tail)  &  1.06(0.63-1.79)&  0.819   &   - & -      \\ \hline
\end{tabular}
\label{tab:stats} 
\end{table}

\begin{figure*}[htb]
	\centering
	\includegraphics[width=0.85\linewidth]{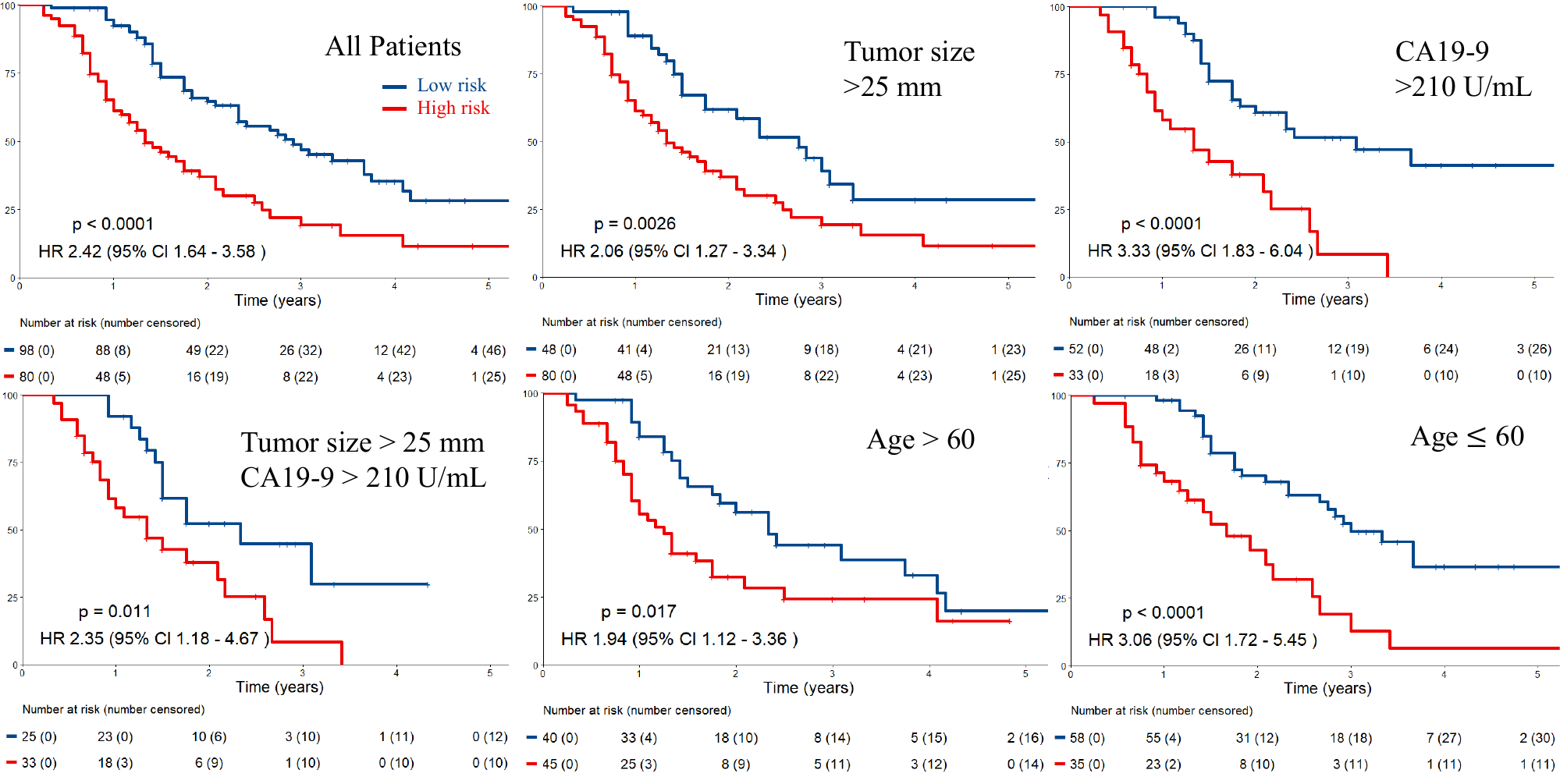}
	\caption{Kaplan-Meier analyses of overall survival according to the proposed signature in all patients in the independent test set (n=178) and subgroups defined by preoperative factors. High risk group indicated by the proposed method is the potential patient group that could benefit from neoadjuvant treatment before surgery.
	}
	\label{fig: kmplot}
\end{figure*}

\noindent\textbf{Neoadjuvant therapy selection.} To demonstrate the added value of our signature as a tool to select patients for neoadjuvant treatment before surgery, we plotted Kaplan-Meier survival curves in \autoref{fig: kmplot}. We further stratify patients by our signature after grouping them by tumor size and CA19-9, two clinically used preoperative criteria for selection, and also age. Our signature could significantly stratify patients in all cases and those in the high-risk group had worse outcomes and might be considered as potential neoadjuvant treatment candidates (e.g. 33 high-risk patients with larger tumor size and high CA19-9).

\section{Conclusion}

In our paper, we propose a multi-branch transformer-based framework for predicting cancer survival. Our framework includes a texture-aware transformer that captures both local and global information about the PDAC and pancreas. We also introduce a neural distance to calculate a more reasonable distance between PDAC and vessels, which is highly correlated with PDAC survival. We have extensively evaluated and statistically analyzed our proposed method, demonstrating its effectiveness. Furthermore, our model can be combined with established high-risk features to aid in the patient selections who might benefit from neoadjuvant therapy before surgery.

\noindent
\textbf{Acknowledgement}
This work was supported by Alibaba Group through Alibaba Research Intern Program. Bin Dong and Li Zhang was partly supported by NSFC 12090022 and 11831002, and Clinical Medicine Plus X-Young Scholars Project of Peking University PKU2023LCXQ041. Yu Shi was supported by the National Natural Science Foundation of China (No. 82071885).

\bibliography{paper1077}
\bibliographystyle{splncs04}
\newpage
%

\newcolumntype{P}[1]{>{\centering\arraybackslash}p{#1}}

\title{Improved Prognostic Prediction of Pancreatic Cancer Using Multi-Phase CT by Integrating Neural Distance and Texture-Aware Transformer}
\titlerunning{Improved Prognostic Prediction of Pancreatic Cancer}
%
\author{Supplementary Material}
\authorrunning{H. Dong et al.}
%
\institute{}

%
\maketitle              

\begin{figure*}[htb]
\centering
\caption{Five visual examples of images, ground truths and spatial information among tumor and vessels. Sorted in ascending order by survival time.}

\includegraphics[width=0.82\textwidth]{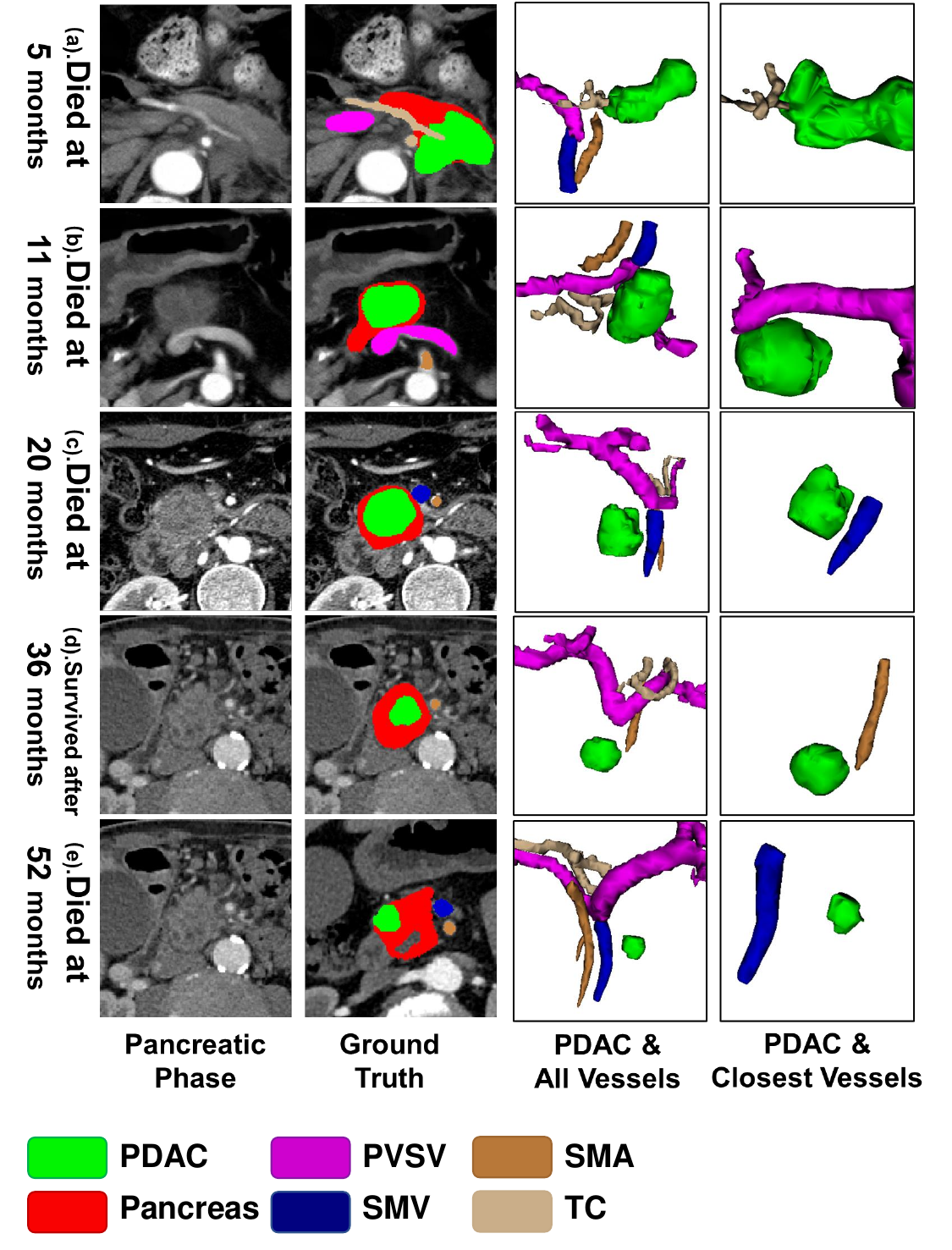}  
\label{fig:4}
\end{figure*}

\noindent\textbf{Details on nnUnet}. 
The multi-phase CT images were directly concatenated as input channels and the pseudo-annotations were generated by combining two teacher models. The student model was trained on both manual and pseudo-annotated multi-phase images, with the teacher models trained for 200 epochs (250 batches per epoch) and the student models trained for 33 epochs (1500 batches per epoch).

\noindent\textbf{More details on our proposed framework}.
The Texture-aware blocks consist of 3x3x3 and 1x1x1 convolution layers, with the former capturing local spatial information and the latter mapping the input tensor to a higher dimensional space. In Fig. 2, the convolution layer highlighted in blue is a single layer, while the yellow block represents the entire CNN block, composed of a convolution layer, a batch normalization layer, and a LeakyReLU activation layer.

\begin{table}[t]
\centering
\setlength{\tabcolsep}{5pt}
\caption{Survival prediction results for DeepCT-PDAC method and our proposed method on the examples in \autoref{fig:4}. Our proposed method, using the texture-aware transformer and neural distance to capture tumor enhancement patterns and tumor-vascular involvement, predicts more reasonable scores on cases with various of survival times. }

\begin{tabular}{c|c|c|c|P{1.5cm}}
\hline
Case  & \thead{survival \\ times} & status & DeepCT-PDAC & Ours  \\ \hline
a & 5  & \text{died} & 0.90 & 0.98 \\ \hline
b & 11 & \text{died} & 0.79 & 0.91 \\ \hline
c & 20 & \text{died} & 0.89 & 0.79 \\ \hline
d & 36 & \text{censored} & 0.65 & 0.48 \\ \hline
e & 52 & \text{died} & 0.60 & 0.08 \\ \hline
\end{tabular}\label{tab:sample}
\end{table}

\begin{table}[hbt]
\centering
\setlength{\tabcolsep}{5pt}
\caption{Ablation study for point size $K$ in nerual distance. We used nested 5-fold cross-validation to evaluate the performance of the model at different values of $K$, and found that $K=32$ achieved the best results.}

\begin{tabular}{|c|c|}
\hline
K  & C-index  \\ \hline
16 & 0.648 $\pm$ 0.021   \\ \hline
32 & 0.656 $\pm$ 0.017  \\ \hline
64 & 0.655 $\pm$ 0.023  \\ \hline
128& 0.653 $\pm$ 0.016 \\ \hline
\end{tabular}\label{tab:Kab}
\end{table}

\vspace{-5mm}


%
%
%

\end{document}